\documentclass{article}

\usepackage{arxiv}
\usepackage{graphicx}

\usepackage[utf8]{inputenc} 
\usepackage[T1]{fontenc}    
\usepackage{hyperref}       
\usepackage{url}            
\usepackage{booktabs}       
\usepackage{amsfonts}       
\usepackage{nicefrac}       
\usepackage{microtype}      
\usepackage{lipsum}

\title{Attaining Real-Time Super-Resolution for Microscopic Images using GAN}

\author{
  Vibhu Bhatia \thanks{Author is currently a part of Boston University. Majority of the work was done while author was in Netaji Subhas University Of Technology.} \\
  Department of Biological Sciences and Engineering\\
  Netaji Subhas University Of Technology\\
  Dwarka, New Delhi, 110078, India \\
   \And
 Yatender Kumar \\
  Department of Biological Sciences and Engineering\\
  Netaji Subhas University Of Technology\\
  Dwarka, New Delhi, 110078, India \\
  \texttt{yatender.kumar@nsit.ac.in} \\
}

\begin{document}
\maketitle

\begin{abstract}
In the last few years, several deep learning models, especially Generative Adversarial Networks have received a lot of attention for the task of Single Image Super-Resolution (SISR). These methods focus on building an end-to-end framework, which produce a high resolution(SR) image from a given low resolution(LR) image in a single step to achieve state-of-the-art performance. This paper focuses on improving an existing deep-learning based method to perform Super-Resolution Microscopy in real-time using a standard GPU. For this, we first propose a tiling strategy, which takes advantage of parallelism provided by a GPU to speed up the network training process. Further, we suggest simple changes to the architecture of the generator and the discriminator of SRGAN. Subsequently, We compare the quality and the running time for the outputs produced by our model, opening its applications in different areas like low-end benchtop and even mobile microscopy. Finally, we explore the possibility of the trained network to produce High-Resolution HR outputs for different domains.
\end{abstract}


\section{Introduction}
Single Image Super-Resolution (SISR) is a widely studied problem in computer vision which aims to generate a High Resolution (HR) Image from a given Low Resolution (LR) counterpart \cite{zomet, Glasner2009}. It is a highly challenging task, but it has been receiving increasing attention from the community over the past few years \cite{7780576,8014885,kim2016accurate,kim2016deeply}. Many of the approaches have been based on using Mean Squared Error (MSE) as a loss function, which, gives a high Peak-Signal-to-Noise-Ratio (PSNR) but fails to capture perceptually relevant details. Recent work by Ledig et al. \cite{ledig2017photo} and Wang et al. \cite{wang2018esrgan} employ the use of Generative Adversarial Networks (GAN) for tackling the problem of super resolution, which have outperformed all other methods and produced state-of-the-art results. Imaging on conventional benchtop microscopes produces images, which are typically limited to their resolution captured by the imaging software regardless of the magnification and the numerical aperture of the lens used. As a result, a compromise must exist between the quality of the image obtained and the field of view (FOV). With the advent of Super-Resolution Microscopy techniques like STEM microscopy \cite{browning2000scanning} and STORM microscopy \cite{rust2006sub}, it is now possible to access even the inner workings of the cell and obtain very high-quality data. However, such methods often rely on very sophisticated setups, matching regions of different images obtained and considerable post-processing to produce high-resolution images. Therefore, these methods require trained personnel to operate and are limited to only a few laboratories around the world. In this paper, we tackle the problem of super resolution on microscopic images by suggesting improvements in the model of SRGAN to primarily improve quality of the images obtained while simultaneously being fast enough to produce results in real time.

\subsection{Related Work}

Prediction-based methods were among the first methods to tackle the Single Image Super-Resolution (SISR). While filtering approaches like Linear, Bicubic or Lanczos filtering\cite{keys1981cubic,duchon1979lanczos}, can be very fast, they oversimplify the SISR problem and usually yield solutions with overly smooth textures. Methods that focused explicitly on edge-preservation had also been proposed \cite{li2001new}. Sun et al. \cite{sun2008image} proposed a gradient profile before describing the shape of gradients. Some of the more powerful approaches rely on training data and tend to establish a relation between the LR and the HR image. Early work on these lines was carried out by Freeman et al. \cite{freeman2002example}. Glasner et al. \cite{Glasner2009} exploited the property of multiple patch occurrences within the image to drive the SR. Some methods also focussed on building an internal database of patches to construct images \cite{freedman2011image}. Other methods which concentrated on Neighbor Embedding \cite{chang2004super}, Sparse Coding \cite{yang2008image,zeyde2010single} and Random Forests \cite{schulter2015fast} were also used for image reconstruction and super-resolution. Methods which uses clustering \cite{yang2013fast} in the latent space of patch were also been used to tackle this problem.

Among Deep Learning (DL) based approaches Dong et al. \cite{dong2014learning} used bicubic interpolation to upscale an input image and trained a three-layer deep fully convolutional network end-to-end to achieve state-of-the-art SR performance. Similarly, works by Johnson et al. \cite{johnson2016perceptual} and Bruna et al. \cite{bruna2015super}, rely on a loss function which relies on perceptual similarity to produce HR images. Ledig et al. \cite{ledig2017photo} proposed the use of residual blocks and content loss to train a GAN based network to perform resolution known as SRGAN. Following this, Lim et al. \cite{8014885} employed deeper networks without Batch Normalization (BN) layers. Wang et al. \cite{wang2018esrgan} introduced a new network ESRGAN, which used residual-in-residual dense blocks (RRDB), in addition to removing the BN layers and proposed a relativistic discriminator which provided the current state-of-the-art results.

Convolutional Neural Nets (CNN) \cite{krizhevsky2012imagenet} have shown amazing capabilities for localization of objects without any need of supervised training process. Bazzani et al. \cite{bazzani2016self} used CNN for self-taught object localization on trained networks. Oquab et al. \cite{oquab2014learning} found Global Max Pooling layer could obtain a point on the boundary of the object. This concept was further taken by Zhou et al. \cite{zhou2016learning} who used Global Average Pooling Layer to find out regions in an image containing an object. Ghosh et al. \cite{ghosh2018adgap} proposed an input size-independent network capable of localizing the position of instances of classes. Wang et al. \cite{wang2019deep} presented a deep learning-enabled super-resolution network which used a U-net \cite{ronneberger2015u} like architecture as the generator as opposed to the SRGAN based generator and could convert images obtained using a 10X resolution to  those obtained with 20X ones. Zhou et al. \cite{zhou2018deep} proposed a convolutional neural net (CNN) based approach for super-resolution localization microscopy while Saurabh et al. \cite{gawande2018generative}  used an SRGAN based approach on low-resolution fluorescence microscopy images used in the Cyto 2017 challenge. Zhang et al.  \cite{zhang2019high} proposed a registration free GAN based model for providing resolution enhancement in conventional microscopy.

\section{Methods and Experiments}

\subsection{Data generation and tiling strategy}
The Human Protein Atlas Image Classification Challenge \cite{kaggle} was a competition hosted in Kaggle, which encourages people to develop models which can classify proteins in microscopic images. For our purpose, this dataset provided the most suitable images for our purpose. These images were gathered using immunofluorescence confocal microscopy. The training dataset used contains a total of 124,288 image given in 4 channels: red, blue, green and yellow, where red: Antibody-based staining of Microtubules, blue: DAPI staining of the Nucleus, green: Protein localizations, yellow: Endoplasmic reticulum. The images were generated by a script which produces a random number between 1 and 4 per image, to make them in the order given in table 1. This scheme ensured a diversity of images obtained for training. A total of 10,000 such images shown in Figure 1 were used for this experiment. 
\setlength{\tabcolsep}{4pt}
\begin{table}
\begin{center}
\caption{
Data Generation strategy used
}
\label{table:headings}
\begin{tabular}{lll}
\hline\noalign{\smallskip}
Number & Color Channels & Organelles$\qquad$\\
\noalign{\smallskip}
\hline
\noalign{\smallskip}
1 & r & Microtubules\\
2 & r + b & Microtubules and Nucleus\\
3 & r + b + g & Microtubules, Nucleus and Proteins\\
4 & r + b + g + y & Microtubules, Nucleus, Proteins and ER\\
\hline
\end{tabular}
\end{center}
\end{table}
\setlength{\tabcolsep}{1.4pt}

\begin{figure}
\centering
\includegraphics[width=40mm]{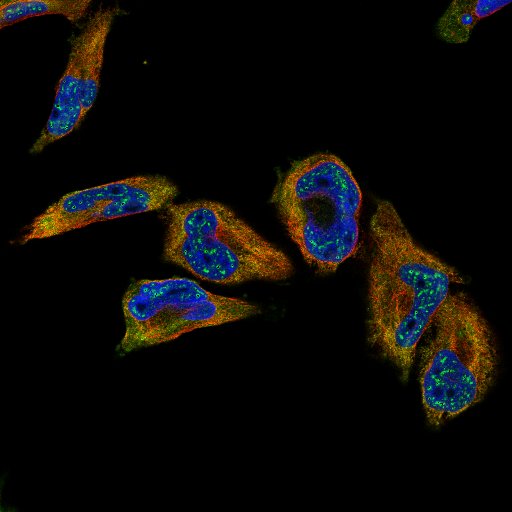}
\includegraphics[width=40mm]{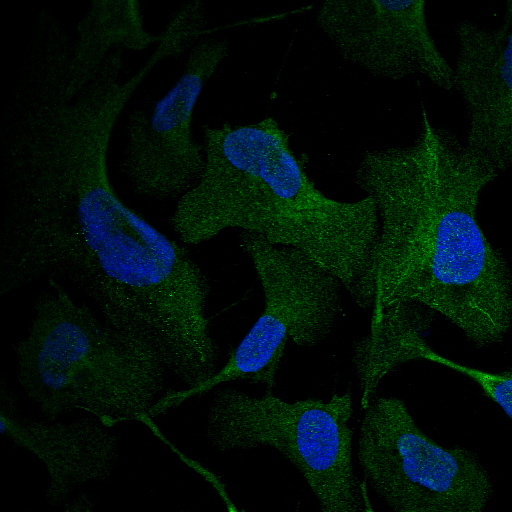}
\includegraphics[width=40mm]{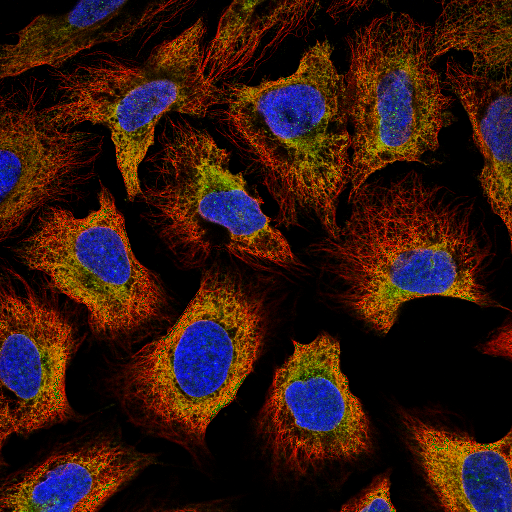}
\caption{Some data points from the dataset used for training the model. Image Source: Human Protein Atlas \cite{kaggle}. }
\label{fig:1}
\end{figure}

The final generated images were of 512 x 512 pixels which were down-sampled 4X using Bicubic Sampling. Since Bicubic sampling produces the lowest quality images, it makes it more difficult for the generator to produce high quality outputs. In order to speed up the image generation process, and to take advantage of the speedup offered by a GPU, both the images were divided into different tiles or patches. Various patch sizes were tested to ensure an optimal tradeoff between an optimal field of view, localization of cells and faster running time for our generator and for our task we found that a tile size of 64x64 pixels for LR image and 256x256 pixels for HR image were the most optimal ones.


\subsection{SRGAN architecture and Improvements}
Ledig et al. \cite{ledig2017photo} proposed the architecture for SRGAN, which we used as our base model, and over which we build up different changes. As we trained our baseline model on the dataset, we started to notice checkerboard-like patterns early in the training phase as shown in Figure 2(a). These patterns even stayed in the image after training for 200 epochs of the entire dataset. We found out this is due to the fact that in the deconvolution layers used for upscaling the image, uneven overlaps may exist if kernel size is not divisible by the stride \cite{odena2016deconvolution}. When this operation is performed over two dimensions like in an image channel, these overlaps get multiplied along the length and width of the image as Figure 2(b). Neural nets typically use more than one of these layers, so it is straightforward for these effects to increase exponentially over multiple layers.
\begin{figure}[h]
\centering
\includegraphics[width=70mm]{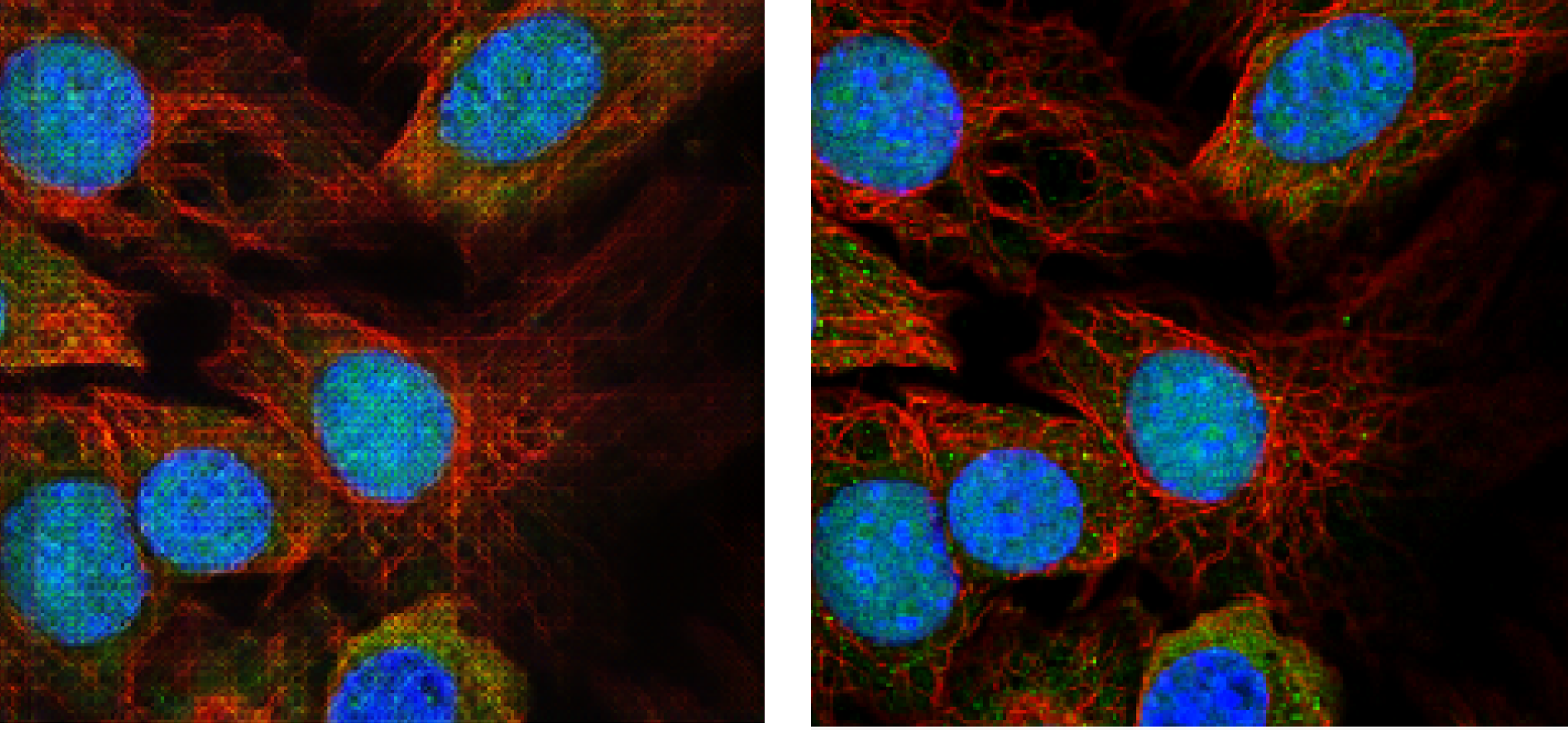}
\includegraphics[width=40mm]{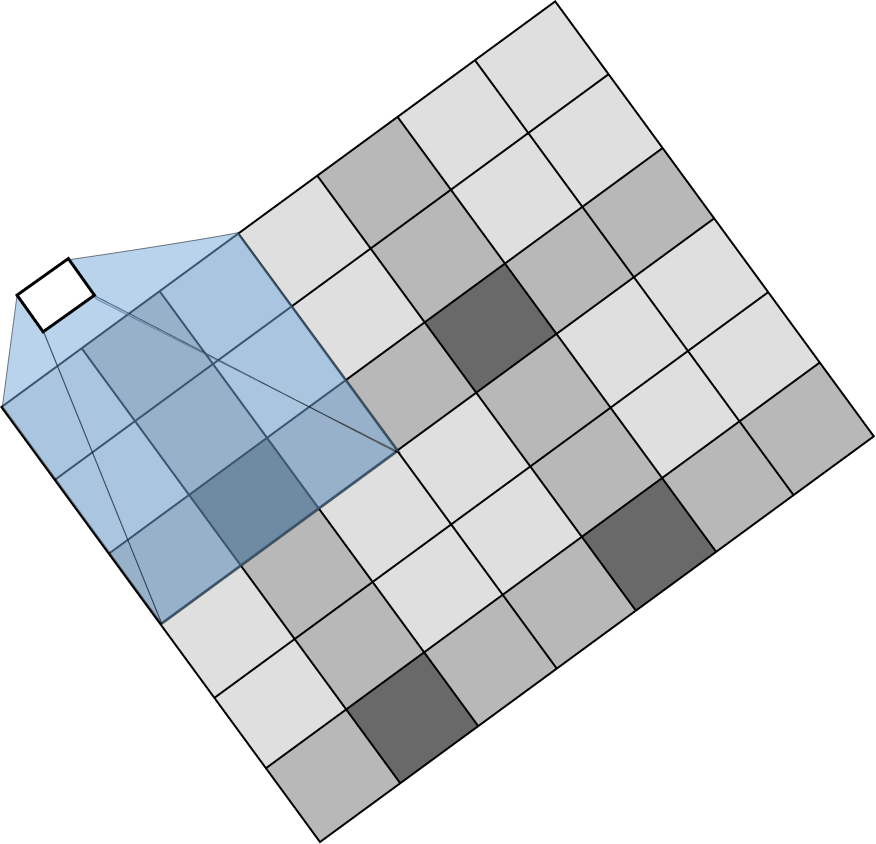}
\caption{a) checkerboard pattern produced on training SRGAN model after the 200th epoch vs. the ground truth image. b) Overlaps produced in standard deconvolution algorithm.}
\label{fig: 2}
\end{figure}

To overcome this problem, a subpixel convolution layer \cite{7780576} is proposed, which works on using a kernel size divisible by the strides. This method is used in SRGAN, but even then, it is possible for these artifacts to occur. The fact that these artifacts remain even in subpixel convolutional layers suggest that even complex network struggle to learn to avoid these artifacts. Different approaches to solve this problem have been implemented by Dong et al. \cite{dong2015image} and Odena et al. \cite{odena2016deconvolution}. Instead of applying deconvolution layers to the image,the task is separated into 2 different processes : upscaling followed by convolution, which seemed to generate better results as shown in Figure 3. For the upscaling part, different methods like Nearest neighbor and bilinear interpolation were tested. We obtained our best results using nearest neighbor approach.Although there was some difficulty in obtaining good results with bilinear interpolation, we believe that a specific initialization of weights could produce good results with bilinear interpolation too.
\begin{figure}
\centering
\includegraphics[width=70mm]{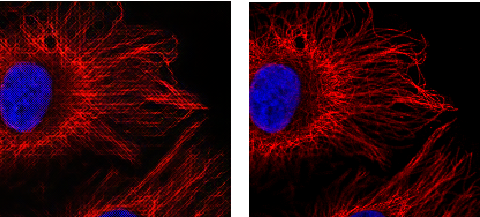} 
\caption{Output generated by original SRGAN vs our modified GAN for different images after 130 epochs.}
\label{fig: 3}
\end{figure}

Another change that we implemented was the removal of Batch Normalization (BN) layers from the residual blocks. As demonstrated by EDSR \cite{8014885} and confirmed by our experiments, removing the Batch Normalization layer seems to reduce the computational time, and by normalizing the mean and variance of each residual block layer, BN layers tend to clip the flexibility available to a generator. Figure 4 shows the comparison between the conventional and the improved residual block.

\begin{figure}[ht!]
\centering
\includegraphics[width=60mm]{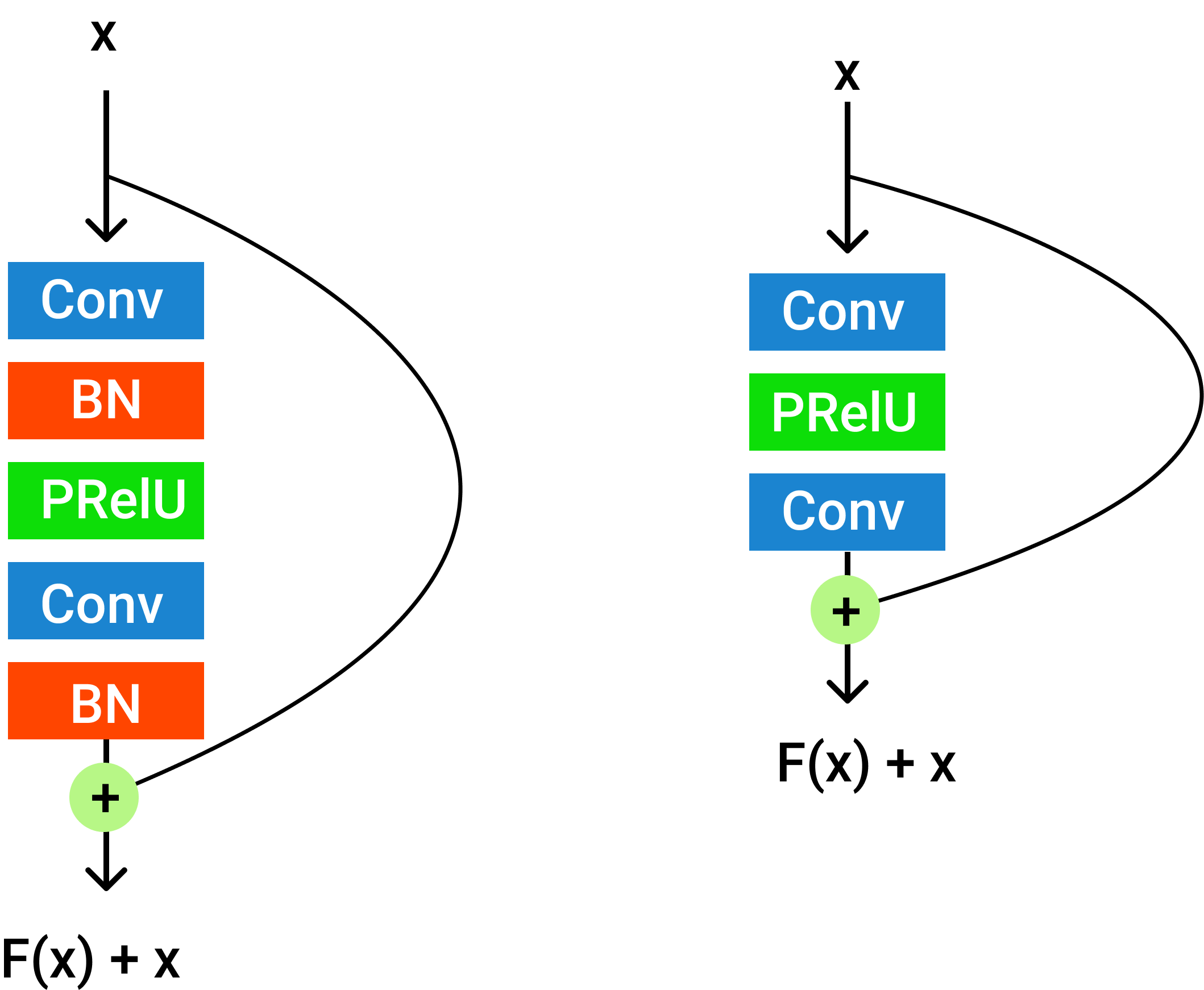} 
\caption{
Conventional residual block in a SRGAN vs Modified residual block.}
\label{fig: 4}
\end{figure}

For the discriminator, we propose the addition of a Global Average Pooling (GAP) layer to the output of the last convolutional block. A GAP layer acts by averaging the pixel values in a feature channel to give one final output which is then directly connected to the final classification layer. GAP layers have excellent localization abilities \cite{zhou2016learning} and also act as a regularizer and help to prevent overfitting while reducing the total number of parameters \cite{Lin2014NetworkIN}. These reasons made using a GAP layer as an option worth exploring in place of the flatten layer proposed in the original architecture.

\subsection{Loss Function and Label Smoothing}
Perceptual loss function was used by ledig et. al for the training of SRGAN\cite{johnson2016perceptual}. It consists of 2 components - the adversarial loss and the content loss. The adversarial loss trains the Generator to produce more solutions which resemble that of the dataset. The content loss, on the other hand, compares high level feature representations of the  images using a separate pre-trained VGG network and enables the network to produce results which are perceptually similar \cite{bruna2015super}.

For the discriminator, we propose using label smoothing given by salimans et. al. \cite{Salimans}. We specifically use one sided label smoothing i.e. Only for discriminator. Target labels for real images are 1, and for fake images are 0 which were replaced with a random number between 0.8 and 1.2 for real and 0 and 0.2 for fake labels. This helps in reducing confidence in real and fake class for discriminator and improves the stability of discriminator during our training.

\subsection{Training schedule and Inference process}
The modified SRGAN network which consists of a discriminator and a generator is used for training on the dataset of confocal microscopy images from section 2.1. The images were first divided into tiles, down sampled using bicubic filter using PIL image library in python to obtain noisy LR images, and then fed into the generator. The generated SR images and the HR images were then fed to the discriminator, which outputs the probability of the generated image as an actual HR image. The training process is shown in Figure 5.
\begin{figure}
\centering
\includegraphics[width=150mm]{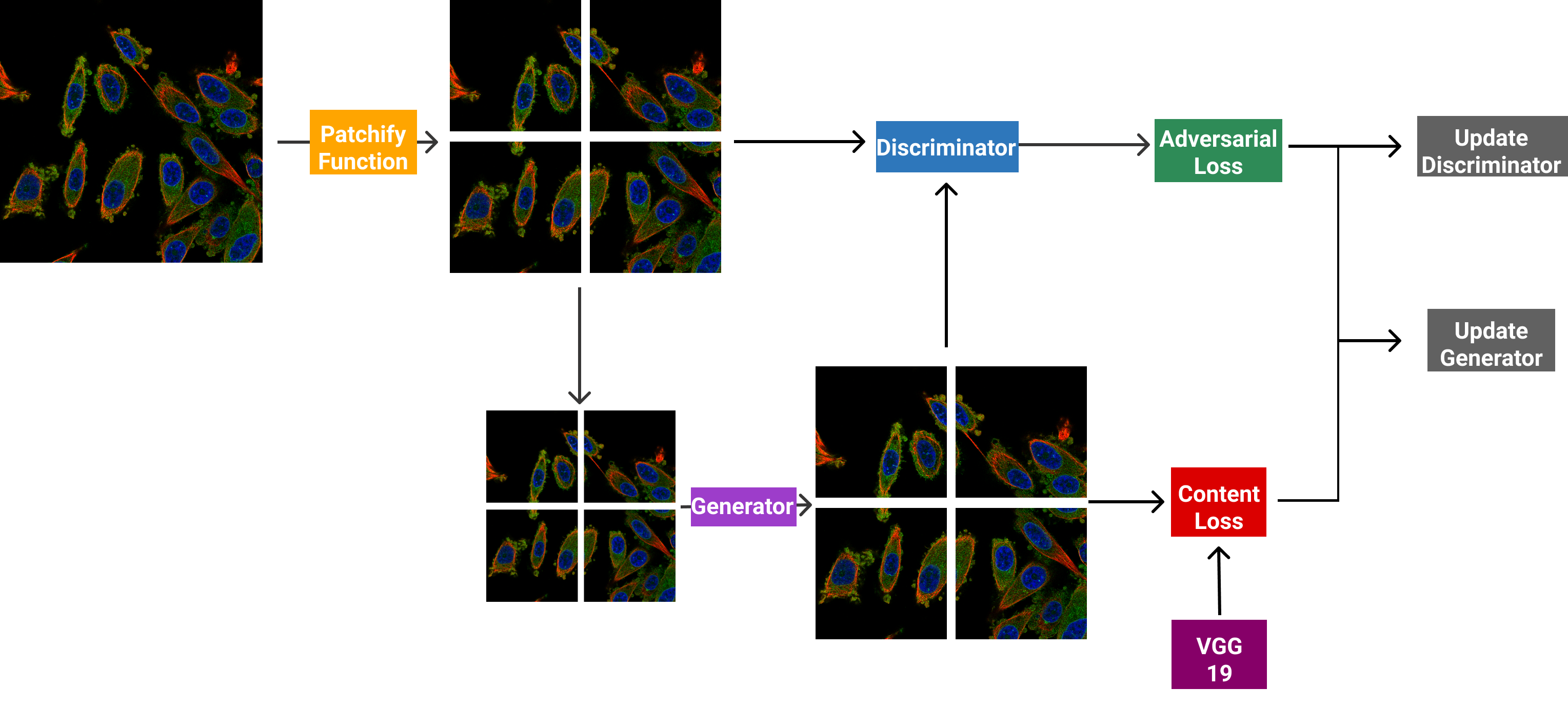} 
\caption{
Training Process for our modified GAN.}
\label{fig:mod GAN}
\end{figure}

All the experiments were performed by downsampling the HR images 4 times via bicubic interpolation to obtain noisy Low-Resolution Images. Each HR image sized 512 x 512 pixels, was divided into tiles of 256 x 256, thereby resulting in a total of 4 images per HR image. The amount of such images used in one iteration depends on the memory of the GPU used. In our experiments, we tried different patch sizes, such as that of 64 x 64, 128 x 128, 256 x 256 and 512 x 512 pixels per image. We found that an enormous patch size does not improve the finer details but increases the overall quality of the image and increases the computational time per image. On the other hand, a smaller patch size like 128 x 128 improves the finer details and saves a lot in computational time but would limit the Field of view of the microscope. Therefore, a middle ground was established between the finer details and the receptive field of the image, which was found to be 256 x 256 pixels. We trained our network on a Nvidia Tesla K80 GPU on a sample of 10,000 images obtained from the Human Protein Atlas Image Classification Challenge \cite{kaggle}. All the models were trained for 200 epochs with each epoch containing 1000 iterations, which corresponds to a total of 2 x \begin{math}10^5 \end{math} iterations. We used learning rate scheduler which used value of 1e-4 for the first half of the optimization process and then reduced to 1e-5 for the next half, while using Adam optimizer with beta1 = 0.9 and beta2 = 0.99.The trained model was thereafter integrated into a web-based portal was developed for the inference process. \footnote{Source Code for training can be found at: \\ \url{https://github.com/vibss2397/Real-Time-Super-Resolution-For-Microscopic-Images-Using-GAN}.}
\subsection{Image Similarity Measures}
Different metrics for testing the quality of the image obtained have been used. However, there is no direct method for measuring visual similarity or improvement. In our experiments, we majorly found PSNR, SSIM and MOS to be most relevant. PSNR is normally used as a quality ratio for measuring the reconstruction quality from a loss compression. It compares the quality of reconstruction between different images. In our case, the SR Image generated by the generator is compared to the original image to measure the loss and subsequently compared it. However, PSNR is inherently biased to favor overly smooth results. SSIM is another measure of perceptual similarity between 2 images but unlike PSNR it also takes high level features such as edges into account. Therefore, in order to improve SSIM, algorithms have to preserve higher order features in addition to removing the noise from the reconstructed images. MOS or Mean Opinion Score is a metric which is used to measure the quality of experience of an Image, Audio or Video. We believe that it is also essential since it directly takes how a human visually processes and compares these images.
\section{Experimental Results}
\subsection{Qualitative Results}
For evaluating the results of the Generator, we compared the results of the generator predicted tiles with the nearest neighbor and the ground truth image at the beginning and the end of the training process. These results are shown in Figure 6. As we can see, the generator is initially trying to figure out the colors and generates weak upscaled versions which contain artifacts and are blurry, but over time, the quality of the image attempts to improve, and by the 200th epoch the generated output is very similar to the ground truth image.

\begin{figure}[h!]
\centering
\includegraphics[width=100mm]{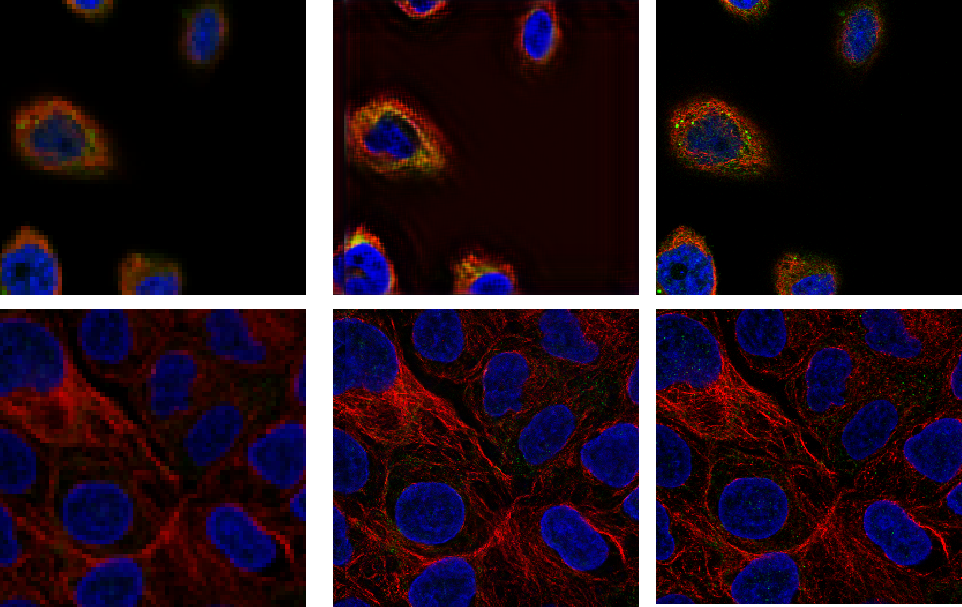} 
\caption{Image produced by upscaling LR Image using Nearest Neighbor Interpolation, the network output, and the Ground Truth Image at the end of 0 and 200 epochs.}
\label{fig: 6}
\end{figure}

Further, Figure 7 compares the results obtained on images from the validation set with different modifications we applied to the SRGAN network and used to train our model. The changes to the baseline were stacked on top of one another. The final architecture, which is SRGAN with Nearest Neighbor Interpolation in the up-sampling block, and no Batch Normalization, produces results which are of better quality than the other methods used. We also tested the ability of out generator to provide cross domain results. (see Appendix , section A.)
\begin{figure}[h!]
\centering
\includegraphics[width=120mm]{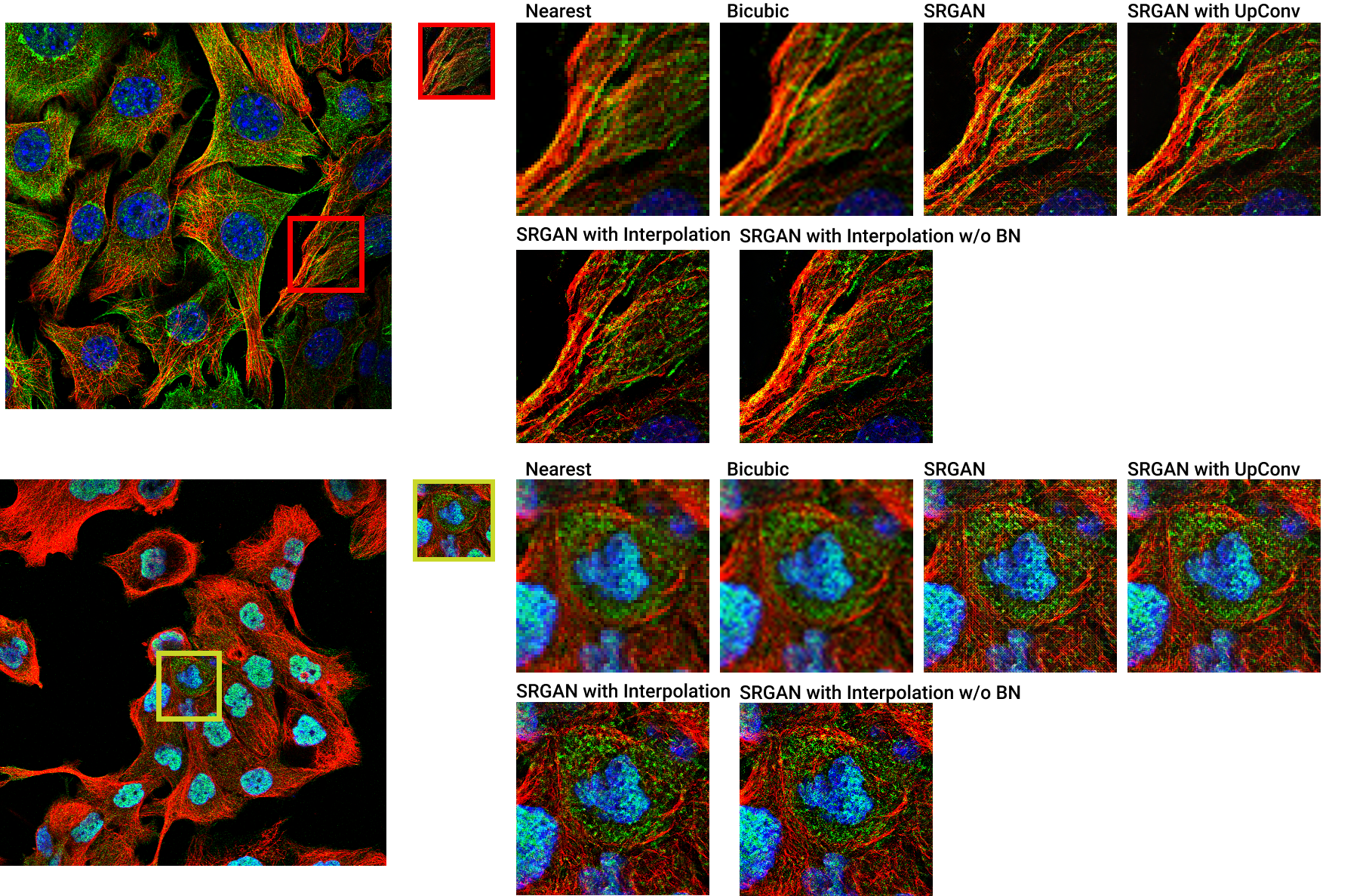} 
\caption{Image created by upscaling LR Image using Nearest Neighbor Interpolation, Bicubic interpolation, SRGAN and different modifications proposed for the generator and the discriminator.}
\label{fig: 7}
\end{figure}

\subsection{Quantitative Results}
For calculating the MOS, we conducted a survey of around 98 anonymous people who were provided with a set of 36 images, which were generated from the outputs of the modified generator, original SRGAN, bicubic approach, nearest neighbor approach, and ground truth image. The images were created and presented in random order, and the raters were asked to rate the images on a scale of 1-5. The results are summarized in Figure 8.  As we can see, most of the scores for bicubic and Nearest Neighbor approach lie in the range of 1, 2 and 3. For SRGAN, they lie between 3 and 4, but for our Modified SRGAN and Ground Truth, they lie in the field of 4 and 5, which suggests a substantial similarity in visual quality between images generated by our generator and the ground truth image. These results are also shown in figure 8.

\begin{figure}
\centering
\includegraphics[width=70mm]{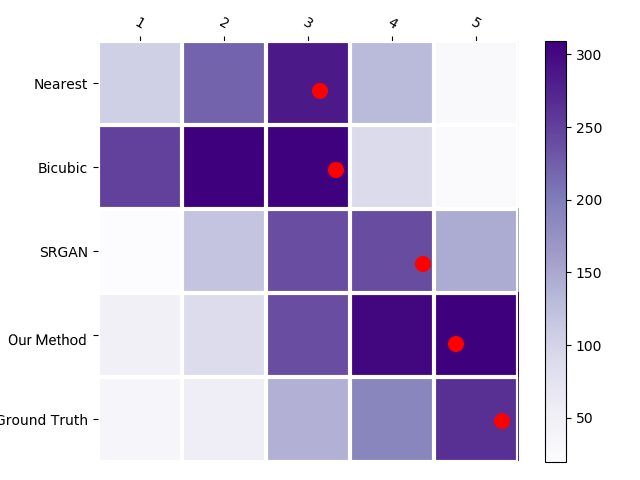} 
\caption{A heat map based on the scores of people rating the images generated by different strategies based on visual quality on a scale of 1-5.}
\label{fig: 8}
\end{figure}

Furthermore, PSNR and SSIM calculated for the subset of the test images, and these scores along with MOS, are given in Table 2. As we can see that for the test set, SRGAN performed better than all the other methods in tests such as PSNR, SSIM, and MOS and was very close to even the ground truth Image in the MOS.

\setlength{\tabcolsep}{4pt}
\begin{table}
\begin{center}
\caption{
Benchmarking and comparison of some image similarity measures for different upscaling methods.
}
\label{table:2}
\begin{tabular}{llll}
\hline\noalign{\smallskip}
Generated Image $\qquad$& PSNR & SSIM & MOS\\
\noalign{\smallskip}
\hline
\noalign{\smallskip}
SRGAN  & 32.585 & 0.815 & 3.9\\
\textbf{Our Method} & \textbf{33.873} & \textbf{0.832} & \textbf{4.2}\\
Nearest Neighbor  & 27.556 & 0.771 & 2.7\\
Bicubic  & 28.498 & 0.810 & 2.9\\
Ground Truth  & $\infty$ & 1 & 4.8\\

\hline
\end{tabular}
\end{center}
\end{table}
\setlength{\tabcolsep}{1.4pt}
\subsection{Evaluation of Running Time for Modified SRGAN}
We evaluated the Running Time of our model for different images in the test set. The running time recorded for two tasks – 1) for upscaling a specific selected patch and the time for dividing an image into patches, and converting them into their SR forms and, 2) combining them back using an unpatch function into the 4x image obtained. All the experiments were performed on a Nvidia Tesla k80 GPU. Our model was able to perform SISR on a 64 x 64 patch to produce a 256 x 256 image in an average time of 0.051 seconds. In case of converting the entire image, we took the time for the entire process, which includes splitting the image into patches, performing predictions on them and finally stitching the images back to one final image. The 512 x 512 image was downscaled to 128 x 128 using Bicubic sampling and time taken to generate the resulting 512 x 512 image was recorded whose average was found out to be 0.14 seconds.

We further test the capabilities of our network to perform the inference in real-time for a given video. We tested the ability of our network to perform super-resolution for a given ROI in the frame of the video. A custom script was developed to select a region from each frame, and only the time taken to perform super-resolution and processing of data was recorded. The video that we used is publicly available as Light Sheet Fluorescence Microscope of a Zebrafish Heart \cite{zebrafish}. Our model took 0.05 seconds to convert a single frame. The results are presented in Table 3 highlighting the fact that our model can produce results in a frame rate of 24.2. Another thing to notice is that removal of the Batch Normalization layer increases the fps of the video generated. These results are consistent with our initial assumptions that Batch Normalization results in the addition of extra parameters to our model and with its removal, our model produces results faster.

\setlength{\tabcolsep}{4pt}
\begin{table}
\begin{center}
\caption{
Evaluation of runtime for obtaining SR images with different modifications to the network for Single Patch(sec), Whole Image(sec) and for a single 64 x 64 patch in the video(FPS).
}
\label{table:3}
\begin{tabular}{llll}
\hline\noalign{\smallskip}
Method & Single Patch & Whole Image & Video Patch\\
\noalign{\smallskip}
\hline
\noalign{\smallskip}
SRGAN  & 0.067 & 0.30 & 19.9\\
SRGAN with upconv, & 0.069 & 0.33 & 20.1\\
SRGAN with interpolation  & 0.089 & 0.34 & 19.1\\
SRGAN w/o BN & 0.041 & 0.12 & 25.7\\
\textbf{Our Method} & \textbf{0.051} & \textbf{0.14} & \textbf{24.2}\\
Nearest Neighbor & 0.0009 & 0.00042 & 99.9\\
Bicubic Interpolation & 0.00016 & 0.0035 & 134.5\\
\hline
\end{tabular}
\end{center}
\end{table}
\setlength{\tabcolsep}{1.4pt}
\subsection{Cross Domain results provided by modified SRGAN}
We test the capabilities of our pre-trained generator to produce results for images obtained in other domains. The ability of a GAN network to produce SR images for cases where the data lies outside the domain of its training set demonstrates its flexibility, and we believe, is something that should be investigated in the future for the development of a more generalized super-resolution system. Figure 11 shows our modified GAN producing SR outputs for images from different datasets like set 14, GFP based fluorescence staining and cross-section of a mouse brain \textbf{[43]} (Figure 10). However, the network struggles to produce clear SR outputs for the regions containing bright or white colors (Figure 9). We believe that this is due to the lack of white color in the images used in a dataset and the use of a diverse dataset which includes more colors like imaging of tissues and in pathology if included in the training dataset might produce better outputs in such cases.
\begin{figure}[h]
\centering
\includegraphics[width=120mm]{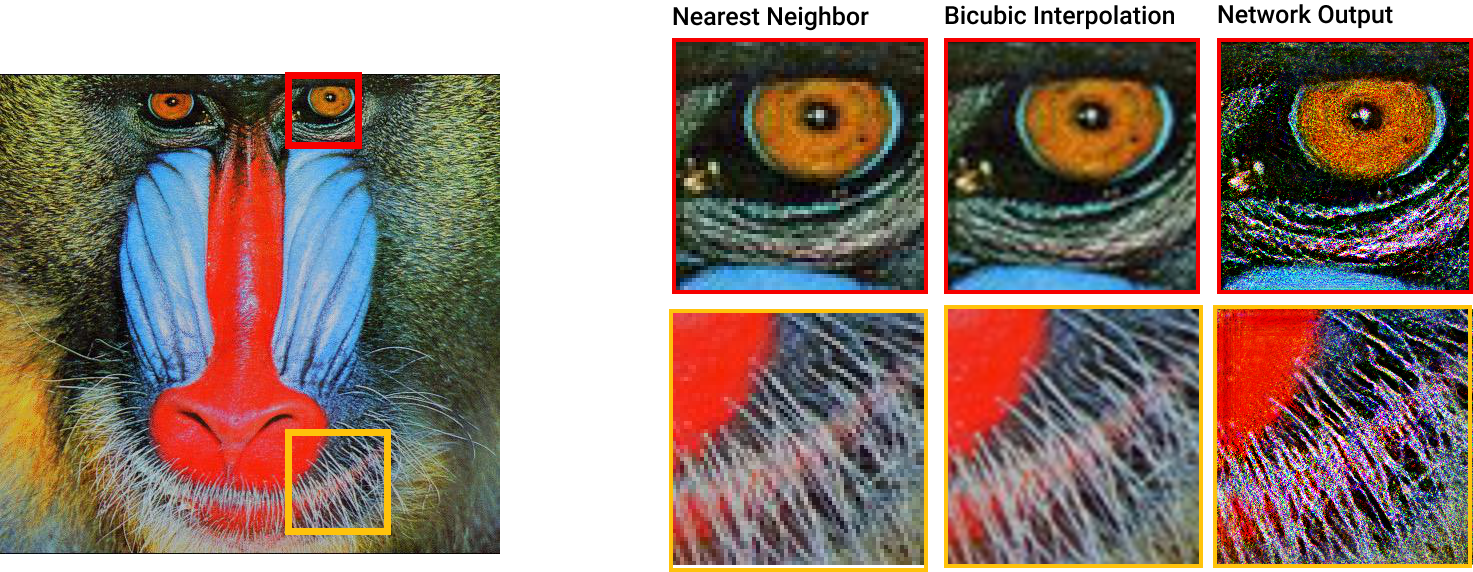} 
\caption{Comparison of cross-domain results for different parts of the image of baboon available in Set 14 dataset.}
\label{fig: 9}
\end{figure}

\begin{figure}[h]
\centering
\includegraphics[width=80mm]{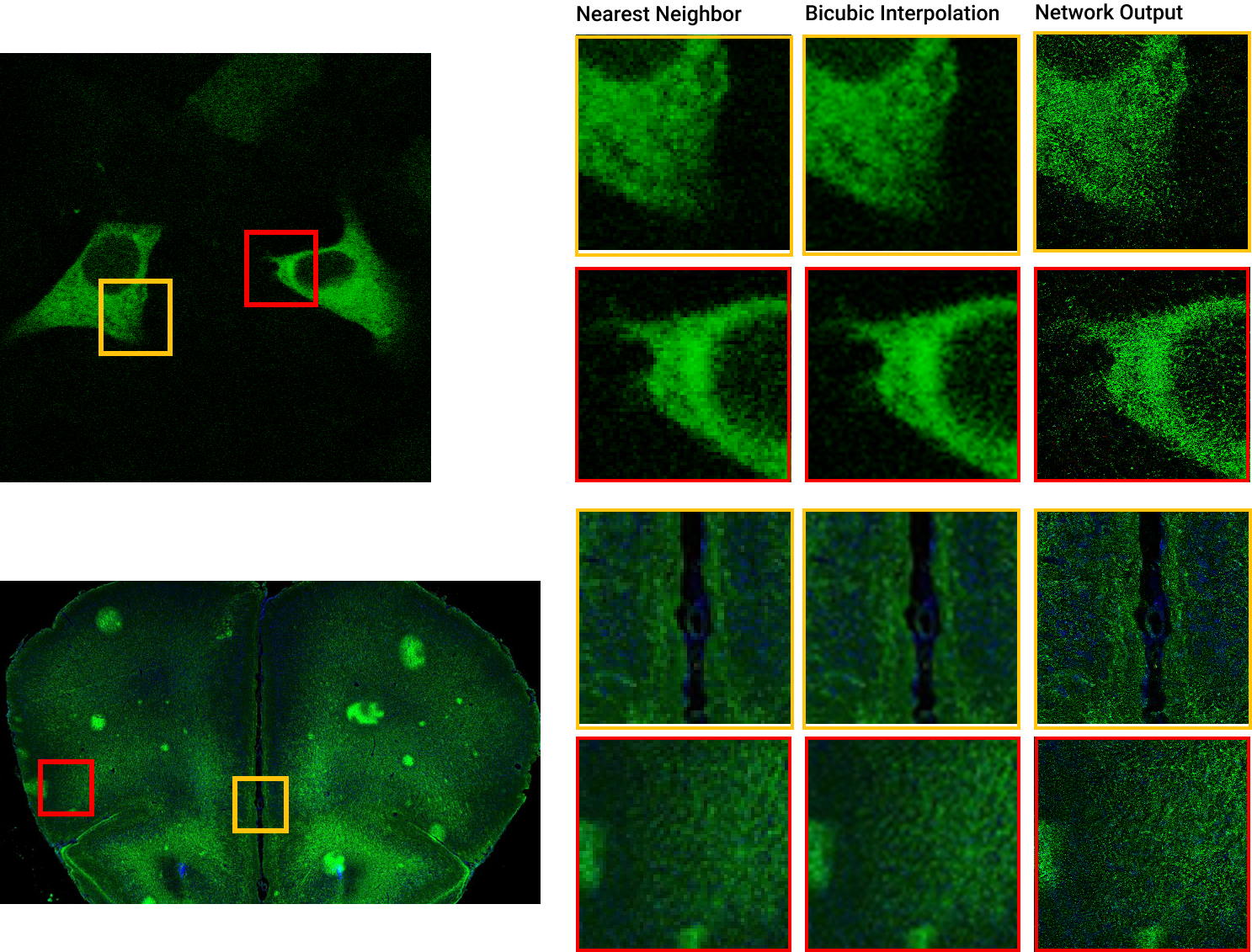} 
\caption{Comparison of cross-domain results for different parts of the fluorescent image of cells and mouse brain (Image Source: Human Protein Atlas).}
\label{fig: 10}
\end{figure}
\section{Conclusion}
This paper demonstrates some of the problems that modern deep learning-based architectures for SISR, specifically SRGAN face. We propose changes in the up sampling and the residual blocks of the generator and the addition of GAP layer to the discriminator, which showed improved results for super-resolution. We then train this modified SRGAN architecture on the dataset provided by Atlas Human Protein Identification Challenge for its use in enhancing microscopic images. Further, we test the viability of the trained generator in other cases for images in different domains, such as natural images, biomedical histology and pathology images. We also validated our results by performing MOS tests that the network does produce outputs which lie closer to the ground truth image in terms of the overall structure and perceptual quality.

\bibliographystyle{unsrt}  
\bibliography{references}  

\begin{thebibliography}{10}

\bibitem{zomet}
Assaf Zomet and Shmuel Peleg.
\newblock Multi-sensor super-resolution.
\newblock In {\em Proceedings of the Sixth IEEE Workshop on Applications of
  Computer Vision}, WACV ’02, page~27, USA, 2002. IEEE Computer Society.

\bibitem{Glasner2009}
Daniel Glasner, Shai Bagon, and Michal Irani.
\newblock Super-resolution from a single image.
\newblock In {\em ICCV}, 2009.

\bibitem{7780576}
W.~{Shi}, J.~{Caballero}, F.~{Huszár}, J.~{Totz}, A.~P. {Aitken}, R.~{Bishop},
  D.~{Rueckert}, and Z.~{Wang}.
\newblock Real-time single image and video super-resolution using an efficient
  sub-pixel convolutional neural network.
\newblock In {\em 2016 IEEE Conference on Computer Vision and Pattern
  Recognition (CVPR)}, pages 1874--1883, 2016.

\bibitem{8014885}
B.~{Lim}, S.~{Son}, H.~{Kim}, S.~{Nah}, and K.~M. {Lee}.
\newblock Enhanced deep residual networks for single image super-resolution.
\newblock In {\em 2017 IEEE Conference on Computer Vision and Pattern
  Recognition Workshops (CVPRW)}, pages 1132--1140, 2017.

\bibitem{kim2016accurate}
Jiwon Kim, Jung Kwon~Lee, and Kyoung Mu~Lee.
\newblock Accurate image super-resolution using very deep convolutional
  networks.
\newblock In {\em Proceedings of the IEEE conference on computer vision and
  pattern recognition}, pages 1646--1654, 2016.

\bibitem{kim2016deeply}
Jiwon Kim, Jung Kwon~Lee, and Kyoung Mu~Lee.
\newblock Deeply-recursive convolutional network for image super-resolution.
\newblock In {\em Proceedings of the IEEE conference on computer vision and
  pattern recognition}, pages 1637--1645, 2016.

\bibitem{ledig2017photo}
Christian Ledig, Lucas Theis, Ferenc Husz{\'a}r, Jose Caballero, Andrew
  Cunningham, Alejandro Acosta, Andrew Aitken, Alykhan Tejani, Johannes Totz,
  Zehan Wang, et~al.
\newblock Photo-realistic single image super-resolution using a generative
  adversarial network.
\newblock In {\em Proceedings of the IEEE conference on computer vision and
  pattern recognition}, pages 4681--4690, 2017.

\bibitem{wang2018esrgan}
Xintao Wang, Ke~Yu, Shixiang Wu, Jinjin Gu, Yihao Liu, Chao Dong, Yu~Qiao, and
  Chen Change~Loy.
\newblock Esrgan: Enhanced super-resolution generative adversarial networks.
\newblock In {\em Proceedings of the European Conference on Computer Vision
  (ECCV)}, pages 0--0, 2018.

\bibitem{browning2000scanning}
ND~Browning, EM~James, K~Kishida, I~Arslan, JP~Buban, JA~Zaborac, SJ~Pennycook,
  Y~Xin, and G~Duscher.
\newblock Scanning transmission electron microscopy: an experimental tool for
  atomic scale interface science.
\newblock {\em Reviews on Advanced Materials Science(Russia)}, 1(1):1--26,
  2000.

\bibitem{rust2006sub}
Michael~J Rust, Mark Bates, and Xiaowei Zhuang.
\newblock Sub-diffraction-limit imaging by stochastic optical reconstruction
  microscopy (storm).
\newblock {\em Nature methods}, 3(10):793--796, 2006.

\bibitem{keys1981cubic}
Robert Keys.
\newblock Cubic convolution interpolation for digital image processing.
\newblock {\em IEEE transactions on acoustics, speech, and signal processing},
  29(6):1153--1160, 1981.

\bibitem{duchon1979lanczos}
Claude~E Duchon.
\newblock Lanczos filtering in one and two dimensions.
\newblock {\em Journal of applied meteorology}, 18(8):1016--1022, 1979.

\bibitem{li2001new}
Xin Li and Michael~T Orchard.
\newblock New edge-directed interpolation.
\newblock {\em IEEE transactions on image processing}, 10(10):1521--1527, 2001.

\bibitem{sun2008image}
Jian Sun, Zongben Xu, and Heung-Yeung Shum.
\newblock Image super-resolution using gradient profile prior.
\newblock In {\em 2008 IEEE Conference on Computer Vision and Pattern
  Recognition}, pages 1--8. IEEE, 2008.

\bibitem{freeman2002example}
William~T Freeman, Thouis~R Jones, and Egon~C Pasztor.
\newblock Example-based super-resolution.
\newblock {\em IEEE Computer graphics and Applications}, 22(2):56--65, 2002.

\bibitem{freedman2011image}
Gilad Freedman and Raanan Fattal.
\newblock Image and video upscaling from local self-examples.
\newblock {\em ACM Transactions on Graphics (TOG)}, 30(2):1--11, 2011.

\bibitem{chang2004super}
Hong Chang, Dit-Yan Yeung, and Yimin Xiong.
\newblock Super-resolution through neighbor embedding.
\newblock In {\em Proceedings of the 2004 IEEE Computer Society Conference on
  Computer Vision and Pattern Recognition, 2004. CVPR 2004.}, volume~1, pages
  I--I. IEEE, 2004.

\bibitem{yang2008image}
Jianchao Yang, John Wright, Thomas Huang, and Yi~Ma.
\newblock Image super-resolution as sparse representation of raw image patches.
\newblock In {\em 2008 IEEE conference on computer vision and pattern
  recognition}, pages 1--8. IEEE, 2008.

\bibitem{zeyde2010single}
Roman Zeyde, Michael Elad, and Matan Protter.
\newblock On single image scale-up using sparse-representations.
\newblock In {\em International conference on curves and surfaces}, pages
  711--730. Springer, 2010.

\bibitem{schulter2015fast}
Samuel Schulter, Christian Leistner, and Horst Bischof.
\newblock Fast and accurate image upscaling with super-resolution forests.
\newblock In {\em Proceedings of the IEEE Conference on Computer Vision and
  Pattern Recognition}, pages 3791--3799, 2015.

\bibitem{yang2013fast}
Chih-Yuan Yang and Ming-Hsuan Yang.
\newblock Fast direct super-resolution by simple functions.
\newblock In {\em Proceedings of the IEEE international conference on computer
  vision}, pages 561--568, 2013.

\bibitem{dong2014learning}
Chao Dong, Chen~Change Loy, Kaiming He, and Xiaoou Tang.
\newblock Learning a deep convolutional network for image super-resolution.
\newblock In {\em European conference on computer vision}, pages 184--199.
  Springer, 2014.

\bibitem{johnson2016perceptual}
Justin Johnson, Alexandre Alahi, and Li~Fei-Fei.
\newblock Perceptual losses for real-time style transfer and super-resolution.
\newblock In {\em European conference on computer vision}, pages 694--711.
  Springer, 2016.

\bibitem{bruna2015super}
Joan Bruna, Pablo Sprechmann, and Yann LeCun.
\newblock Super-resolution with deep convolutional sufficient statistics.
\newblock {\em arXiv preprint arXiv:1511.05666}, 2015.

\bibitem{krizhevsky2012imagenet}
Alex Krizhevsky, Ilya Sutskever, and Geoffrey~E Hinton.
\newblock Imagenet classification with deep convolutional neural networks.
\newblock In {\em Advances in neural information processing systems}, pages
  1097--1105, 2012.

\bibitem{bazzani2016self}
Loris Bazzani, Alessandra Bergamo, Dragomir Anguelov, and Lorenzo Torresani.
\newblock Self-taught object localization with deep networks.
\newblock In {\em 2016 IEEE winter conference on applications of computer
  vision (WACV)}, pages 1--9. IEEE, 2016.

\bibitem{oquab2014learning}
Maxime Oquab, Leon Bottou, Ivan Laptev, and Josef Sivic.
\newblock Learning and transferring mid-level image representations using
  convolutional neural networks.
\newblock In {\em Proceedings of the IEEE conference on computer vision and
  pattern recognition}, pages 1717--1724, 2014.

\bibitem{zhou2016learning}
Bolei Zhou, Aditya Khosla, Agata Lapedriza, Aude Oliva, and Antonio Torralba.
\newblock Learning deep features for discriminative localization.
\newblock In {\em Proceedings of the IEEE conference on computer vision and
  pattern recognition}, pages 2921--2929, 2016.

\bibitem{ghosh2018adgap}
Arna Ghosh, Biswarup Bhattacharya, and Somnath Basu~Roy Chowdhury.
\newblock Adgap: Advanced global average pooling.
\newblock In {\em Thirty-Second AAAI Conference on Artificial Intelligence},
  2018.

\bibitem{wang2019deep}
Hongda Wang, Yair Rivenson, Yiyin Jin, Zhensong Wei, Ronald Gao, Harun
  G{\"u}nayd{\i}n, Laurent~A Bentolila, Comert Kural, and Aydogan Ozcan.
\newblock Deep learning enables cross-modality super-resolution in fluorescence
  microscopy.
\newblock {\em Nature methods}, 16(1):103--110, 2019.

\bibitem{ronneberger2015u}
Olaf Ronneberger, Philipp Fischer, and Thomas Brox.
\newblock U-net: Convolutional networks for biomedical image segmentation.
\newblock In {\em International Conference on Medical image computing and
  computer-assisted intervention}, pages 234--241. Springer, 2015.

\bibitem{zhou2018deep}
Tianyang Zhou, Jianwen Luo, and Xin Liu.
\newblock Deep learning for super-resolution localization microscopy.
\newblock In {\em Optics in Health Care and Biomedical Optics VIII}, volume
  10820, page 1082023. International Society for Optics and Photonics, 2018.

\bibitem{gawande2018generative}
Saurabh Gawande.
\newblock Generative adversarial networks for single image super resolution in
  microscopy images, 2018.

\bibitem{zhang2019high}
Hao Zhang, Chunyu Fang, Xinlin Xie, Yicong Yang, Wei Mei, Di~Jin, and Peng Fei.
\newblock High-throughput, high-resolution deep learning microscopy based on
  registration-free generative adversarial network.
\newblock {\em Biomedical Optics Express}, 10(3):1044--1063, 2019.

\bibitem{kaggle}
Human~Protein Atlas.
\newblock {\em Human Protein Atlas Image Classification | Kaggle.}, October 3,
  2018(accessed, 30-November-2019).

\bibitem{odena2016deconvolution}
Augustus Odena, Vincent Dumoulin, and Chris Olah.
\newblock Deconvolution and checkerboard artifacts.
\newblock {\em Distill}, 1(10):e3, 2016.

\bibitem{dong2015image}
Chao Dong, Chen~Change Loy, Kaiming He, and Xiaoou Tang.
\newblock Image super-resolution using deep convolutional networks.
\newblock {\em IEEE transactions on pattern analysis and machine intelligence},
  38(2):295--307, 2015.

\bibitem{Lin2014NetworkIN}
Min Lin, Qiang Chen, and Shuicheng Yan.
\newblock Network in network.
\newblock {\em CoRR}, abs/1312.4400, 2014.

\bibitem{Salimans}
Tim Salimans, Ian Goodfellow, Wojciech Zaremba, Vicki Cheung, Alec Radford,
  Xi~Chen, and Xi~Chen.
\newblock Improved techniques for training gans.
\newblock In D.~D. Lee, M.~Sugiyama, U.~V. Luxburg, I.~Guyon, and R.~Garnett,
  editors, {\em Advances in Neural Information Processing Systems 29}, pages
  2234--2242. Curran Associates, Inc., 2016.

\bibitem{zebrafish}
Carl Zeiss.
\newblock {\em Lightsheet Z.1 by Carl Zeiss Microscopy - Zebrafish Heart},
  October 3, 2018(Accessed, 30-Sep-2019).

\end{thebibliography}






\end{document}